# TDACS: an ABAC and Trust-based Dynamic Access Control Scheme in Hadoop

**Min Yang**

Sichuan University, Chengdu, 610065, China

**Abstract:** The era of big data has promoted the vigorous development of many industries, boosting the full potential of holistic data-driven analysis. Hadoop has become the primary choice for mainstream platforms used by stakeholders to process big data. Thereafter, the security of Hadoop platform has arisen tremendous attention worldwide. In this paper, we mainly concentrate on enforcing access control on users to ensure platform security. First, we leverage access proxy integrated with attribute-based access control (ABAC) model to implement front-end authorization, which can fully reflect and cope with the flexible nature of the complex access control process in Hadoop platform, as well as can release back-end resources from complex authorization process through access proxy. Moreover, in order to ensure the fine-granularity of authorization, the access proxy maintains a list composed of trust threshold value provided by each resource according to its importance. The access proxy interacts with the blockchain network to obtain the user's trust evaluation value, which serves as an important basis for dynamic authorization determination. More specifically, blockchain network works together on-chain and off-chain modes. The user's historical behavior data is stored off-chain, and the corresponding hash value is anchored on-chain. Consequently, the user's trust value is evaluated based on his historical behavior stored on the blockchain platform. Meanwhile, the authenticity of user behavior data can be guaranteed, thereby ensuring the reliability of trust assessment results. Our experiment demonstrates that the proposed model can dynamically and flexibly adjust user permissions to ensure the security of the platform, while time and money are consumed within a reasonable range.

## 1 Introduction

Big data has brought enormous political and economic benefit varying from fields of business, politics to education, medical care and etc., which has almost penetrated into every aspect of our lives. According to the survey report on the development of big data in China[1], the core industry of big data was 23.6 billion yuan in 2017, with a growth rate of 40.5%, and the growth rate is expected to remain above 30% from 2018 to 2020. However, with the increasing growth of big data generation, access and utilization, the security and privacy of which has arisen a widely public concern[2]. In January 2017, an extortion event took place in MongoDB database[3]. Hackers used the database configuration vulnerability to make unauthorized access and steal sensitive information. In August 2017, the leakage of top secret information of the Swedish government caused serious political risks[4]. In the middle of March 2018, a data leak happened in Facebook[5], and the abuse of a large number of sensitive information such as voter information related to political orientation caused serious political influence. According to Common Vulnerabilities and Exposures vulnerability list[6], from 2012 to 2018, a total of 19 vulnerabilities were exposed in Hadoop, of which 5 were related to the leakage of sensitive information, accounting for 26.3% of the total, showing a sustained growth trend.

Currently, enterprises, governments, and other organizations have built their own data lakes through big data platforms in succession. Among them, Hadoop is the most popular choice, through which not only stakeholders can store and process a large amount of data, but also share their resources with other parties for the common good, but as Reference[7] pointed out the security mechanism of Hadoop platform still has serious deficiencies and limitations, which was initially designed to be used only in trusted internal networks, with less consideration given to authentication and authorized access of external users, leading to obstacles to resource sharing among multi-platforms. Typically, the widely application of Hadoop cloud platform and the boom of data sharing have forced both research and industry to do a lot of work on secure and dependable access control mechanism so as to maximally protect platform resources. The most common seen security mechanisms in Hadoop encompass Kerberos, access control lists (ACL), log monitoring etc.[8]. In summary, current Hadoop security efforts mostly focus on the following three aspects. a)Authentication，the verification of the user or system identity accessing the system. Hadoop provides a combination of Kerberos and tokens as the primary means of authentication. b)Authorization, a process of specifying access control privileges for user or system according to access control policy[9]. In recent years, the industry and academia at home and abroad have conducted in-depth research on access control technology in the context of big data. So far, researchers mainly focus on the following types: RBAC (role-based Access Control, corresponding to the open source project Apache Sentry), PBAC (policy-based Access Control, the represented open source project is Apache Ranger) and ABAC (attribute-based Access Control, using Apache Knox). c) Audit, there is always the possibility of accidental, unauthorized access or inappropriate access by wrong users causing a security breach[10], therefore, to meet security compliance requirements, as early as the 1980s, Anderson proposed the system log analysis to audit illegal access[11], which is broadly employed to solve security violations in Hadoop platform. d)data encryption, which is out of the scope of this paper.

Notably, the above four factors are usually studied separately, most of achievements solve merely one aspect, which is far from achieving the effect of big data platform security protection. Accordingly, the present study proposes a dynamic authorization scheme embedded with one of the most popular blockchain technology to provide safety protection for Hadoop platform, which can solve a), b)at the same time as well as supporting c) within a certain period of time. The main contributions of this paper can be summarized as follows.

1. The combination of the ABAC model and the access proxy guarantees the flexibility and scalability of authorization. At the same time, in order to further refine permissions, trust access thresholds are set for resources to implement trust-based fine-grained access control.

2. Introduce blockchain technology. The immutability of the blockchain provides authenticity guarantees for the storage of user behavior data, and the chain storage also fits the time-related characteristics of user behavior. It provides a strong support for the accuracy of subsequent behavioral data evaluation as well as supporting auditing.

3. The idea of 'sliding window' is introduced to limit user behavior data to a specific length of time window. In this way, the timeliness of trust evaluation can be ensured. Moreover, expired behavior records will be removed from the window and will be cleared by storage systems, which is in accordance with the law of personal information protection.

This paper is structured as follows. Section 2 reviews related work. Section 3 describes the preliminaries and presents some important definitions. Section 4 explains the proposed architecture. We empirically evaluate our method based on artificial data in Section 5. Section 6 presents some brief concluding remarks and and provides future research directions.

## 2 Background and Related Work

In this section, we briefly review the Hadoop platform access control model from different perspectives, and give a primer on the blockchain technology.

## 2.1 Current Access Control Schemes

There have been various attempts to address security issues of Hadoop platform, both from enhancing identity authentication, as well as from developing fine-grained access control scheme. The original Hadoop had no security mechanism in mind, Hadoop 1.0.0 version came with the Kerberos and token mechanism to guarantee the credibility of user identity[12]. However, Kerberos-based access control mechanism cannot effectively secure platform's resource, the Hadoop cluster can be accessed with a verified token issued by the platform without monitoring and assessing user's operation behavior. Wang et al.[13] aimed at solving over-authorization in the Hadoop cloud platform, which attackers can easily use upper-layer application services to attack the underlying Hadoop system platform and cause data security problems, and hence an identity-based access control scheme was designed, and a third-party broker was used to carry out user identity authentication and permission check to solve the authorization problem. Although it can achieve the minimum permission control of a job, it only makes access control decisions based on the user's identity information and a static permList maintained internally by the platform, and there is no supervision of user behavior after authorization. Liu et al.[14] designed a trust-based access model in Hadoop, in which each user had a trust value, and updated it in real time through the user's behavior record in the cluster, thus dynamically control the user's access to the platform. However, the trust assessment method of this model ignores the timeliness of behavior data, and fails to be resistant to users accumulating credit and then launching attacks. Yang et al.[15] presented a dynamic access control system on user behavior assessment for Hadoop cloud platform. The system exploited CSL algorithm to mine the behavior pattern from the user instruction sequence, and then calculated the user behavior evaluation value through 'sliding window algorithm', thus dynamically control the user's access to the platform according to the value. However, the paper does not explain how the trust value is added to the access control policy and what kind of access control model is used, in addition, the platform only gives refusal or permission answers to user access merely based on trust values, the constraints are too few and the granularity is not fine enough. Reference[16] proposed a SecureDL system in which utilized a data access broker to capture user's requests, then analyzed whether these requests conform to security and privacy policies, and recorded them in case of auditing. Reference[17] pointed out that the Hadoop Ecosystem lacks native techniques for securing sensitive data of users stored and processed on the system and proposed a multi-layer policy-based access control scheme to prevent unauthorized access to cluster resources and imposed access control for data owners. Based on the analysis of Hadoop system logs, Li et al.[18] put forward a personalized access control scheme to track the user's access behavior trajectory and gave different permissions to users with different trust levels, which can effectively protect data security through reasonable authorization.

In conclusion, the Kerberos-based Hadoop authentication mechanism once passes authentication, no longer conducts any supervision on users, and the process of identifying user identity is static, which could result in safety problems, so the dynamic access mechanism based on user trust value becomes a better solution, which solves the problem of user behavior supervision. However, access control depends solely on the user's trust value of user operation behavior, ignoring the context environment and other important properties, which would lead to inaccurate authorization. And most of the solutions are targeted at internal users, the study on temporary access to external users scenario is less. In recent years, with the development of cloud computing, attribute-based access control(ABAC) model is increasingly applied to cloud platform[19]. Chen et al.[20] pointed out that ABAC is well suited to big data scenarios because of its fine-grained access control, autonomous authorization, dynamic access control, and low overhead. Chen et al.[21] developed an attribute-based access control model for Hadoop platform. This model expands the traditional ABAC model to a five-tuple, and the addition of security level attributes increases the flexibility and enables the requirements of fine granularity and dynamic authorization. Actually, the research of attribute-based access control model in big data platform is still in the exploration stage due to the unique characteristics of big data.

## 2.2 Blockchain Technology

Although many scholars now take the user trust value as the main basis for authorization and make the access control model more reasonable[12-14,17]. However, most of research work ignored two important factors when implementing user behavior assessments. First, the authenticity of user behavior data is not considered. For example, a malicious user may increase the trust value by modifying historical behavior, while users with good performance may be tampered with their historical behavior data by malicious users to reduce the trust value. Accordingly, ensuring the authenticity of user behavior data is of great significance for subsequent evaluation. Secondly, user behavior changes over time, and the timeliness of behavior data should be fully considered. All user behaviors are considered to be equally important to the evaluation results, which directly affects the accuracy and rationality of the evaluation results. The 'sliding window model' can be used to calculate the trust value, because it can accord with the characteristics of slow rise, fast fall and time correlation of the trust value[22]. However, 'sliding window model' does not consider how to store time-series behavior data. The advent of the blockchain perfectly solves these issues.

Blockchain technology, as a new type of distributed ledger technology, combines a series of computer technologies such as distributed storage, P2P, consensus mechanisms, and encryption algorithms to bootstrap trust in an untrusted network, without a central trust entity. At the same time, it has the characteristics of immutable data, traceable transactions, tampering any transaction of blockchain is an extremely challenging process due to the harness of a cryptographic data structure and the transparency of data among the entire network nodes. Liang et al.[23] designed a decentralized and trusted cloud data provenance model using blockchain technology. The author says that cloud auditing can only be effective if all operations on the data can be traced reliably, and assured provenance data can help detect security violations within the computing infrastructure. So the proposed system collects data provenance record and publishes to the blockchain network so as to build a time-stamped log of all user operations. Data provenance refers to the origin of data records, and can be effectively used to authorization, validation, and access control policy retrieval[24]. There is no doubt that the use of provenance data in audit is an innovative idea, which provides correct and real data support for auditing. However, the audit can't take proactive precautions and can't intercept and respond to violation operation in real time.

Inspired by the above work, in this study, the user behavior data is stored as the provenance data through the blockchain network for subsequently reliable evaluation, and a trust-based dynamic authorization scheme based on blockchain is proposed. This paper adopts one of the most suitable access control models, which is widely employed by cloud platform, ABAC model, to improve the flexibility and dynamics of access control. Using access proxy, external users and internal users can complete access through the proxy to shield the internal details of the cluster. The system can realize 1)Safe storage 2)reliable trust evaluation 3) Fine-grained dynamic access control.

## 3 Preliminaries and Definitions

In this section, some definitions related to the proposed system will be given. In the Hadoop access control model, the user, resource, operation, environment are formally described by ABAC model. Prior to introducing the details of our proposed method, we present several definitions of important concepts related to this paper.

**Definition 1** The ABAC model is defined as four-tuple $AS = \{U_{as}, R_{as}, OP_{as}, E_{as}\}$, the meaning of each element is given separately in the definition below.

**Definition 2** The set of attributes of the user is represented as $U_{as} = \{U_1, U_2, ..., U_n \mid n \in N^+\}$. The user who tries to access the platform resource is denoted as $U_i = \{Uname, Utype, Au \mid n \in N^+\}$, in which *Uname* represents the user name, *Ugroup* represents the group in which the user is located, and different user groups have different initial trust values. *Au* represents the agent of the user who can complete the access to the resource. What needs to be explained here is that this model presets the set of agent users $Au = \{au_1, au_2, ..., au_n \mid n \in N^+\}$ . According to the principle of least privilege, only one proxy user with minimum privileges is assigned to each user to meet their access needs.

**Definition 3** The set of attributes of the resource is represented as $R_{as} = \{R_1, R_2, ..., R_n \mid n \in N^+\}$ . The platform's resource is denoted as: $R_i = \{Rname, Oname\}$ , in which $Rname$ represents the resource name, $Oname$ represents the service name of HDFS, YARN, Hive, and etc.

**Definition 4** The set of attributes of the operation is denoted as $OP_{as} = \{OP\}$ . The user's operation on the resource is defined as $OP = \{read, write, execute, select, drop...\}$ ,which represents the operational attributes of a resource, including read, write, and execute operations on files, select, delete, and drop operations on database tables, and submit operations on job submissions.

**Definition 5** The collection of environment is denoted as $E_{as} = \{ip_{wl}, Speriod, Threshold\}$ . Where $ip_{wl}$ represents white list of IP address, $ip_{wl} = \{ip_1, ip_2, ..., ip_n \mid n \in N^+\}$ . $Speriod$ denotes the time period during which access to the service is allowed $Speriod = \{speriod_1, speriod_2, ..., speriod_n \mid n \in N^+\}$ . $Threshold$ represents a set of threshold values for the number of accesses to a resource, $Threshold = \{N_1, N_2, ..., N_m \mid m \in N^+\}$ .

**Definition 6** The access control decision is based on the policy set, which is defined as $PS = \{Ps, Constraints\}$ . $Ps$ is a collection of policy-related attribute variables group composed of user attributes, resource attributes, and access operations. $Ps = \{X_1, X_2, ..., X_n \mid n \in N^+\}$ . $X_i = \{U_{as}, R_{as}, OP_{as}, status\}$ , in which $status = \{allow, deny\}$ . for example, $X_1 = \{U_1, R_1, write, allow\}$ denotes that, $U_1$ allows 'write' operations on $R_1$ , $Constraints = \{constraint_1, constraint_2, ..., constraint_n\}$ , which represents a set of constraints on the action and environment attributes in a user access request. In this study, $constraint_i = \{C(count, Threshold), T(period, Speriod), IP(ip, ip_{wl})\}$ , where access times, access time and IP address are restricted.

**a）** $C(count, Threshold)$ : Represents the comparison result between the current access times $count$ of a user and the number of accessible services $Threshold$ . The result is represented by C, as shown in formula 1. When the current access times of the user do not reach the accessibility threshold, $count$ increases by 1 ; otherwise, assign $C$ to -1 and the access will be denied.

$$C = \begin{cases} count \leq Threshold \leftrightarrow count++ \\ count > Threshold \Leftrightarrow -1 \end{cases} \tag{1}$$

**b）** $T(tc, Speriod)$ : To determine whether the current access time of the user is within the given access period, formula (2) is used to represent, where $tc$ represents the time period during which a user accesses a resource, $Speriod$ represents the permitted access period. When the user's current access time is within the given access period, $T$ is assigned to $True$ , indicating that access is allowed. Otherwise, $T$ is assigned to $False$ to indicate that access is not allowed.

$$T = \begin{cases} tc \in Speriod \Leftrightarrow True \\ tc \notin Speriod \Leftrightarrow False \end{cases} \tag{2}$$

**c）** $IP(ip, ip_{wl})$ : Indicates whether the user's $ip$ is within the white list $ip_{wl}$ , and is expressed by formula (3). If the $ip$ used by the current user is included in the $ip_{wl}$ , the result is $True$ , indicating that the user is accessible; otherwise, returns $False$ , indicating that the user is not accessible.

$$IP = \begin{cases} ip \in ip_{wl} \Leftrightarrow True \\ ip \notin ip_{wl} \Leftrightarrow False \end{cases} \tag{3}$$

**Definition 7** The weight distribution is inversely related to the time of the behavior data stored by the node.

$$w_i \propto \frac{1}{period} \tag{4}$$

Suppose that in a certain period of time, the number of successful user access requests is $x$ times, and the number of failed access times is $y$ times. The sigmoid function is exploited to calculate the user's trust value. It is worth noting that, the trust value function operates on the rationale that the 'successful access' can be regarded as the user performing well on the platform, and the 'failed access' can be regarded as the user doing evil on the platform (According to Fig. 2, the system flowchart shows that the cause of failure may be token theft, unauthorized access, low trust access etc.). In addition, we can add a penalty factor $\beta$ ( $\beta > 1$ ) to the 'failure' variable. Accordingly, the user's behavior of failed access will rapidly decrease the trust value, which is more in accordance with the actual scenario.

$$ts_i = \frac{1}{1 + e^{-(x - \beta y)}} \tag{5}$$

The main idea behind exploiting this function is for the following reasons:
1. The range is [0,1], thus, the user trust value is normalized to facilitate subsequent processing;
2. The trust value output by this function increases as the number of successes increase, and decreases with the increase of the number of failures, which conforms to the rule of changing the user's trust value;
3. The function commences by a cautious start through an initial slow increase, and drops rapidly under the action of $\beta$, which is consistent with the characteristic of 'slow rise and fast fall' of a user trust value.

## 4 System Design

This section elaborates on the proposed dynamic authorization scheme. In a nutshell, the goal of the system is to build a secure access control environment, where the grant of access right is only related to the reliability of the user identity, as well as the dynamic context information, such as the real-time trust value based on historical behavior, the network environment when initialing request, and etc. In order to achieve this goal, we separate our architecture into two parts, one is called 'ABAC-based authorization', which mainly realized by access proxy, the other is called 'trust-based authorization', mainly built on blockchain network. The overview of our system is illustrated in Fig. 1.

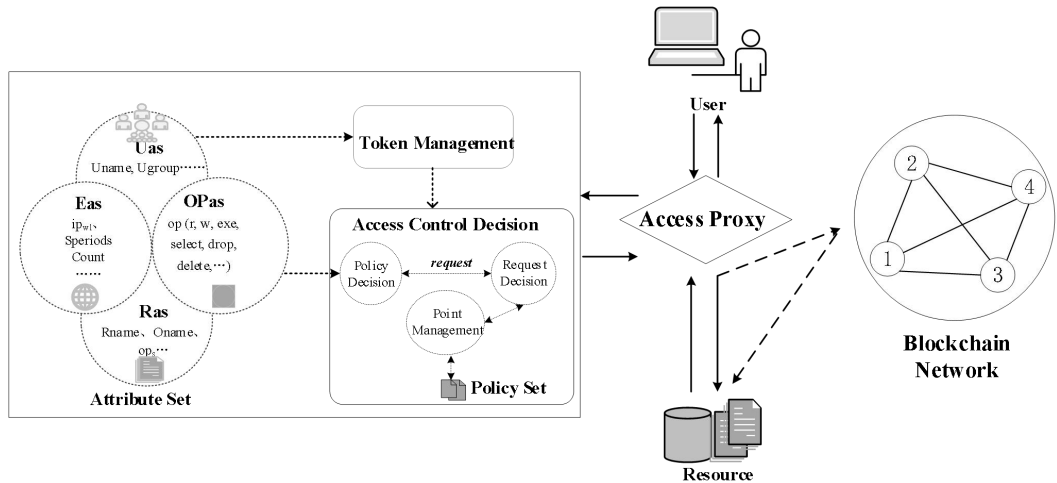

**Fig. 1.** High-level overview of the system architecture

To illustrate, consider the following example: a user who tries to access a certain resource of the Hadoop platform, as the user logs in, a token with access restrictions is generated by Token Management (TM) module and sent to the user, which subsequently acts as credentials for access control decision. Thereafter, the user initiates an access request along with the token, and the Access Control Decision

(ACD) module requires to verify the token implemented by TM, if valid, ACD makes an access control decision based on the policy set, if successfully match, passes the request to the access proxy, then the proxy assigns an agent to the user for handling subsequent interactions. Note that: at this time, the agent cannot successfully obtain the resource, instead, access proxy (AP) initiates a request to the blockchain network (BN) to validate the authenticity of the user historical behavior and then conducts trust value computation, and then compares the value with the accessibility threshold of the resource(reffered as resList in this paper) requested by the user, and subsequently determines whether the user's request is rejected or forwarded to Hadoop service. The specific access process is shown in Fig. 2.

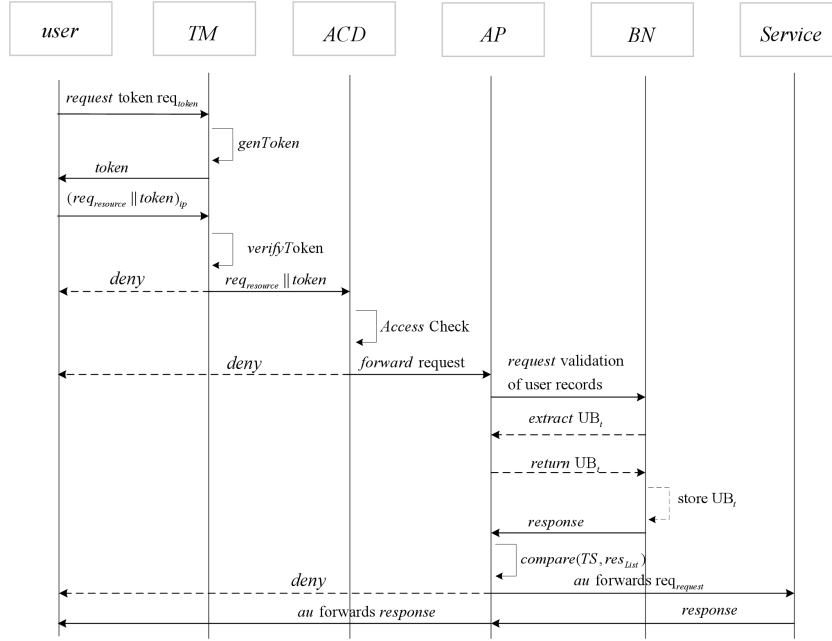

**Fig. 2.** Model operation flow chart

As shown in Tab. 1, we describe the symbols and their corresponding meanings involved in our proposed architecture.

**Tab. 1** Symbols and meanings

| Symbol | Description |
| --- | --- |
| TM | Token Management module |
| ACD | Access Control Decision |
| AP | Access Proxy |
| BN | Blockchain Network |
| Service | Hadoop Service |
| genToken | Token generation |
| verifyToken | Token verification |
| AccessCheck | Access Control Check algorithm |
| TS | user's trust score |
| $UB_T$ | user's historical behavior trajectory |
| ‖ | Message concatenation operation |

In the following, we divide the components and symbols involved in the above process into two parts, that is, ABAC-based and trust-based access control, to explain in detail.

### 4.1 ABAC-based Authorization

This part mainly consists of four critical components.

**User**: who tries to request some resources which may be useful and necessary, users can be internal users of the platform or external users.

**TM**: token management module includes two functions, that is, token generation and token verification. token generation is responsible for generating token for user according to username, password. The token contains constraint fields such as user ip white list, access service period set *Speriod*, access service number set *Threshold*. *IP* white list can ensure the legitimacy of users. *Speriod* and *Threshold* constraint fields ensure the temporary nature of the token, and at the same time do not need to worry about the issue of permission revocation. After the token expires, it will automatically become invalid. Furthermore, it also ensures the security of the token. Token verification is to verify whether the token is stolen or abused by inappropriate users.

**ACD**: contains three sub-modules, which are Request Decision Point (RDP), Policy Management Point (PMP) and Policy Decision Point (PDP). RDP is responsible for verifying the validity of the token and transforming the user requests into attribute-related variable groups based on Attribute Set. PMP mainly provides operations for generating, querying, updating and deleting policies in the policy set *PS*. The function of PDP is to implement policy matching and forward decision results.

**AP**: a centralized policy enforcement coordinated with ACD module, to handle user's access control permissions. In addition, in this study, AP also maintains resList provided by back-end resources to further refine the user access control, realizing the dynamic adjustment of access rights based on user trust value.

The working process of ABAC is described as follows.

**Step1**. Request token. The user requests a token from TM by showing his username, password and ip address, $req_{token} = \{Uname, password, ip\}$.

Step2. Generate token. Upon receiving $req_{token}$ from the user, TM authenticates *Uname* and password contained in the user request, and then obtains the user-related property set from AS, user-related ip address white list $ip_{wl}$, permitted access time *Speriod* and permitted access times *Threshold* according to *Oname*. a token is generated and sent to the user in an encrypted way. $token = Enc\{Uname, Speriods, ip_{wl}, Threshold\}$.

**Step3**. Request resource. After receiving the token, the user initiates a resource request along with the token to ACD. $req_{resource} \| token = \{Uname, Oname, OP, ip\} \| Enc\{Uname, periods, ip_{wl}, t hs\}$.

**Step4**. Verify token. ACD sends the received message to the TM module for token verification. TM carries out the following performance upon receiving. 1) Decrypt the token 2) Obtain the user's ip address from the user's request, and compare it with $ip_{wl}$ included in the token to check if the user of the token is the true owner. If is, TM sends a response to ACD, otherwise, sends 'deny' to user.

**Step5**. Access control decision. Upon receiving the confirmation message, Algorithm1 is employed to implement access control check. If it passes the access control check, ACD forwards the user's access request to AP, otherwise, it returns 'deny' to the user.

---

**Algorithm 1** Access Control Check

**Input:** $req_{token}$, $token$, $req_{resource}$, $AS$, $P_s$

**Output:** *permit*, *deny*;

   1. Get *Uname,Oname,Rname* from $req_{resource}$;

   2. R=find( $AS.R_{as}$,*Rname*);

   3. *periods*=*token*.get('Speriod');

4. Get period from *periods* by R.*Oname*;

5. ths=*token*.get('Threshold');

6. Get th from ths by R.*Oname*;

7. Get *Uname*'s count from Database;

8. **if** T(CurrentTime, period)==True **then** && C(count, th)==1

9.    Database=count++;

10. **end if**

11. Match $req_{resource}$ with $P_s$;

12. **if** match **then**

13.    forward $req_{resource}$ to AP;

14.    **return** 'permit';

15. **else**

16.    **return** 'deny';

17. **end if**

---

After authorization through ABAC, we have emphasized earlier that users cannot immediately obtain a response from the service. Instead, AP derives the trust value of the user by evaluating the user's historical behavior and realizes dynamic and fine-grained authorization based on trust value. Next, we will detail the trust-based authorization.

### 4.2 Trust-based Authorization

In Reference [25], Luca pointed out combining coarse-grained, centralized authorization at the access proxy with fine-grained authorization at the back end provides the best of both worlds, inspired by this, apart from configuring some generic access policies in access proxy, fine granular access control manipulated by the user trust value is added. More specifically, trust value of each resource is set (resList), access to a certain resource must be greater than the threshold. The combination of front-end centralized authorization and resource-level fine-grained access control schemes can achieve more flexible and dynamic authorization, as well as avoid over-authorization. Prior to talking about the back-end fine-grained authorization communication process, the storage and evaluation of user behavior data are firstly introduced.

### 4.2.1 Data Storage

User behavior data has the characteristics of large volume, dynamic nature, and chronological feature. Good data management methods and secure storage modes can lay a good foundation for subsequent evaluation and analysis. Reference [22] employed the cloud service provider to detect user operations in real time (such as the creation, copying, and sharing of cloud files) and uploaded these behavior records to the blockchain network to ensure the authenticity of the source data, and thus providing a reliable source for subsequent audit work. Reference [26] considered from the perspective of secure data sharing and auditable access control, split the IoT data stream into data blocks, and the blocks are linked in a hash manner. Each block contains a pointer to the previous block hash pointers.

Inspired by [22][26], we exploit the blockchain technology, a decentralized database to realize secure and reliable storage. Since the evaluation of behavior data is meaningful only when it is based on a large number of historical behaviors, it is unreasonable to obtain the user trust value from a single behavior record. Therefore, in order to facilitate the subsequent evaluation, we do not store the behavior data per record, but instead store the behavior data within a period of time，maybe by day, week or month and etc.. The operation mode of blockchain network is the collaboration of on-chain and off-chain.

Off-chain, the traditional trusted information system is used to store the original data. On-chain, only a hash pointer of the behavioral data is stored to ensure data immutability.

**Behavioral records use case.** For each user's behavior trajectory, we extract the RecordID, subject, object, operations, time, and flag (flag=1 represents a successful access, flag=0 means a failure access) from AP and store them within a certain period in off-chain database. Before evaluating user behavior, these original behavioral records are validated through the blockchain network to check if the records have been modified. After validation, updating the off-chain database.

**Tab. 2** Structure of single behavioral record

| RecordID | Uname | Rname | OP | tc | flag | Blockhash | TxHash | Validation |
|---|---|---|---|---|---|---|---|---|
| 1 | A | R1 | w | 2020-1-1 9:00 | 1 | hash1 | txHash1 | true |
| 2 | A | R3 | r | 2020-1-1 9:06 | 1 | hash2 | txHash2 | true |
| 3 | B | R1 | delete | 2020-1-1 9:10 | 0 | hash3 | txHash3 | true |
| 4 | B | R2 | insert | 2020-1-1 10:10 | 1 | hash4 | txHash4 | true |
| 5 | B | R2 | copy | 2020-1-1 13:25 | 0 | hash5 | txHash5 | true |
| 6 | A | R3 | drop | 2020-1-2 9:30 | 1 | hash6 | txHash6 | true |
| ... | ... | ... | ... | ... | ... | ... | ... | ... |

*4.2.2 Sliding Window*

The choice of 'sliding window' is motivated by the fact that the user behavior trust evaluation is based on the dynamic update and synthesis of the trust evaluation results of long-term user access, the process of which can be viewed as a sliding window model that slides continuously[27]. In this paper, we set up a sliding window as shown in Fig. 3 on the basis of user behavior data. The effective time span of the trust is $N$, that is, the window length is $N$. When a user initiates a new access request, the leftmost record with the longest time is replaced with a new record in order from left to right. This method allows the weights of the behavioral records stored in the storage units to be changed according to the principle of "far small, near large", that is, data that is far away from current time has a small weight assigned to it, and data that is close to current time has a large weight assigned. This study introduces a sliding window model to store data, on the one hand, to ensure the timeliness of behavioral records, and on the other hand, to comply with the principle of personal information protection, that is, any organization must not permanently store user data.

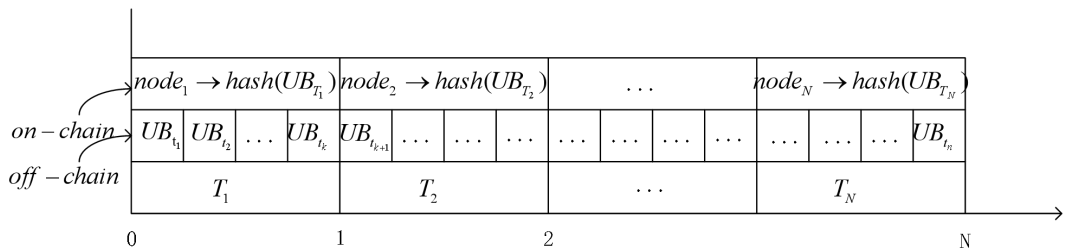

**Fig. 3.** Sliding window-based user behavior data storage

To illustrate, an example is given. Assuming that the sliding window length is $T_N$, and $n$ is the memory space, the behavior data in the window needs to be stored from off-chain in chronological order. Here, for the purpose of simplicity, we assume that the data is divided into 4 parts according to the length of time $T_1$. The sliding window spans a length of $T_4$. The blockchain adopts the on-chain and off-chain collaborative work mode. $L[1] - L[4]$ store the behavioral data in the time period $T_1 - T_4$ separately ($L[k]$ is the storage address of database), denoted as $UB_{T_1} - UB_{T_4}$, and the hash value of the behavior data $hash(UB_{T_1}) - hash(UB_{T_4})$ is stored on-chain from node1 to node4. The weight distributed for user historical records is in accordance with the principle of "far small, near large". Accordingly, the weight vectors of $UB_{T_1} - UB_{T_4}$ can be denoted as $W = [w_1, w_2, w_3, w_4], \sum_{i=1}^{4} w_i = 1, w_4 > w_3 > w_2 > w_1$. When a new behavior record comes, the changing process of off-chain node data and the corresponding weight is shown in Fig. 4.

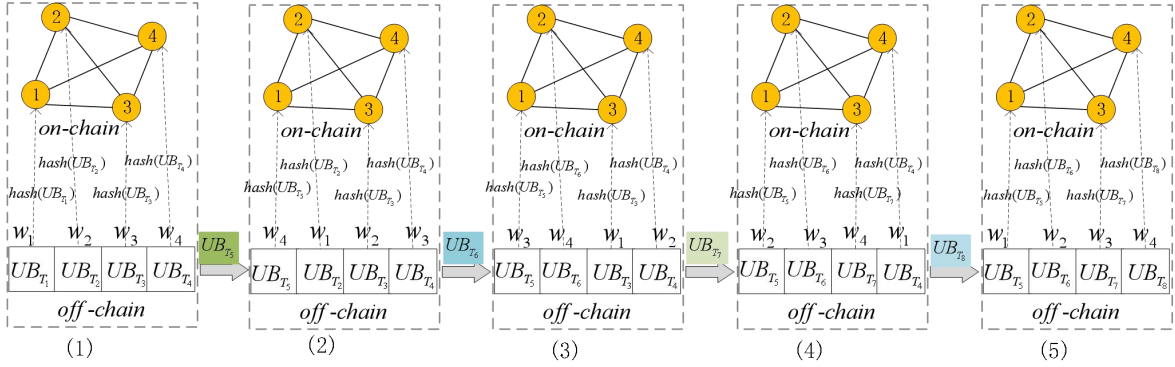

**Fig. 4.** Example of the changing process of data records and weight

(1) In the initial state, the record stored in $L[1] - L[4]$ is $UB_{T_1}, UB_{T_2}, UB_{T_3}, UB_{T_4}$ separately, and the corresponding weight vector is $W = [w_1, w_2, w_3, w_4]$. The hash value of the data is stored on-chain. (2) When a new behavior record $UB_{T_5}$ arrives, the oldest data is pushed out of the window, the data stored from $L[1]$ to $L[4]$ becomes $UB_{T_2}, UB_{T_3}, UB_{T_4}, UB_{T_5}$. (3) When a new access record $UB_{T_6}$ arrives, the conditions of each node both on-chain and off-chain are deduced by analogy and will not be repeated.

### 4.2.3 Trust Evaluation

Before evaluating the trust value of the user, it is necessary to calculate the hash value of the user behavioral data and compare it with the hash value stored in the blockchain network to ensure the authenticity of the user behavior data. If the data has not been tampered with, the weights will be assigned to the behavioral records according to the principle of 'far small, near large', and the trust value in different time periods will be calculated according to formula (5), and then the user's final trust value will be calculated using the following formula.

$$TS = \alpha T_{init} + (1 - \alpha) \sum_{i=1}^{n} w_i \times ts_i \tag{6}$$

In which, $\alpha$ is the influence of the initial trust value on the final trust value, which can be set according to the actual situation. $T_{init}$ refers to the initial trust value, which can be assigned according to the role/position of the user. $n$ is the total storage units, $w_i$ is the weight of the $i$-th storage unit. $ts_i$ is the trust score of the $i$-th storage unit data.

### 4.2.4 Back-end Communication Process

When AP receives the request forwarded by ACD, user's trust evaluation process starts up. Fig. 5 illustrates the concrete communication process, which is divided into two parts, one is data storage: involving the storage of the user's original behavior records and uploading the hash value of the behavior data to the blockchain network. It is worth noting that data storage is a continuous process. Since the user initiates an access request, the user is monitored in real time. Once a new behavior record is generated, it is collected and stored by off-chain storage database. Another part is data processing: including the verification of the authenticity of behavior records, the calculation of user trust score and comparison TS with resource threshold.

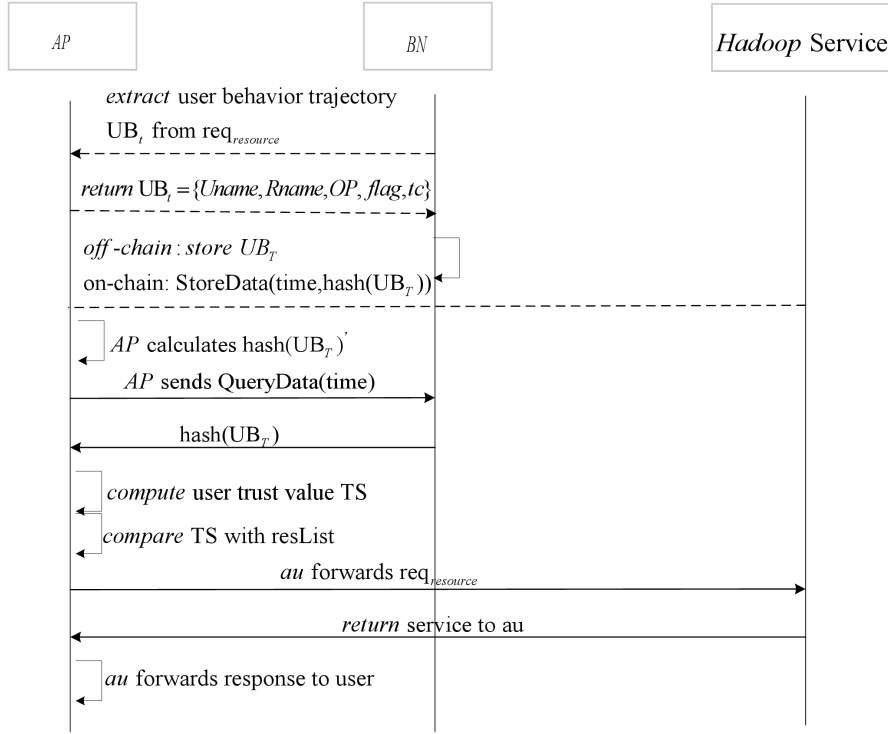

**Fig. 5.** Back-end trust-based access control procedures

## 5 Experiment evaluation

### 5.1 Configuration

Apache Knox[28] offers a single point API gateway to control access to several Hadoop ecosystem services such as Apache Hive, HDFS, YARN etc., to authorize users via pre-configured access strategies. Therefore, in this study, we deploy Knox-0.9.0, Hive-1.2.1000, YARN-2.7.3，Openldap-2.4，Kerberos-1.10.3-10，Ranger-0.6.0 to simulate the access process of a user. Two Docker containers were configured in the experiment. The Hadoop cluster was deployed in a distributed mode, with one container as the master node and the other container as the slave node. The relevant component information deployed on the master and slave nodes is shown in Tab. 4 below.

**Tab. 4** Master and slave node deployment component information

| node | Master and slave node deployment component information |
|---|---|
| master | NameNode、Datanode、SecondNameNode、ResourceManager、NodeManager、Hive Metastore、HiveServer2、Knox Gateway、Openldap-Server、Openldap-Client、Kerberos- |

|  | Server、Ranger Admin、Ranger Usersync |
|---|---|
| slave | DataNode、NodeManager、Openldap-Client、Kerberos-Client、Ranger Tagsync |

## 5.2 Performance

### 5.2.1 Comparison with Knox Native Authorization

We compared the performance difference between adding ABAC module on Knox and Knox native authorization mechanism to verify the impact of increased security on performance.

For the Knox-without-ABAC and Knox-with-ABAC modes, as the amount of data being read increases exponentially, the response time to access the data in HDFS is shown in Tab. 5. Among them, the response time in each mode is the mean value of 10 tests.

**Tab. 5** Comparison of response time between the two modes

| Data size (MB) | 1 | 2 | 4 | 8 | 16 | 32 | 64 | 128 | 512 |
|---|---|---|---|---|---|---|---|---|---|
| without ABAC | 284.6 | 323.8 | 354.4 | 471 | 702 | 1155.1 | 2049.4 | 4223.8 | 9215.8 |
| With ABAC | 292.8 | 334.2 | 363.3 | 479.4 | 714.8 | 1164 | 2062 | 4235.5 | 9224.6 |

With the amount of data read increasing, the difference in response time between the two modes fluctuates between 8-13ms, and there is no obvious growth trend. Therefore, compared with the Knox native authorization mechanism, adding ABAC module on Knox has little impact on performance while enhancing access control for services.

### 5.2.2 The Economic Expenses of Blockchain

Considering that the introduction of blockchain technology in this study is mainly to ensure the authenticity of the behavioral records, to prevent the records from being tampered and affect the user's trust assessment, and it is well known that due to the large number of nodes in the public blockchain, the cost of tampering is much higher than that of the consortium blockchain and private blockchain. Therefore, in this section, we used Ethereum to conduct simulation experiments to verify the economic costs. Specifically, we deployed and ran smart contract on Ethereum to measure the 'Gas' cost of executing a transaction. The smart contract includes two functions:

*StoreData(time,hashValue)* : This function handles data storage request coming from Hadoop platform. As mentioned in Section 4.2, user data records are stored by day/week/month and etc. Accordingly, AP will calculate the hash value *hashValue* of user behavior data during this period, then store it with the timestamp *time* on-chain as the form of key-value pair (*time* | *hashValue*).

*QueryData (time)*: This function tackles with data query request coming from AP or other stakeholders. Every time the user trust assessment is required, AP will send a request to blockchain which includes the function name *QueryData* and the timestamp parameter *time*, then the corresponding *hashValue* will be returned to AP.

Using SHA256 as the hash function, the aforementioned contract was developed on Remix IDE using solidity language, and tested in Ganache[29] which is a personal Ethereum blockchain used to run tests. The 'Gas' cost of contract deployment and the function calls is as shown in Tab. 6.

**Tab. 6** Gas cost of ethereum functions and contract deployment in USD

| The operation of function and contract | Gas | Cost(USD) |
|---|---|---|
| StoreData | 106339 | 0.0013 |
| QueryData | 0 | 0 |
| Contract deployment | 399737 | 0.0047 |

<div align="center">（Gasprice: 1 ether = 20GweiGAS, 1 ether = 237.27 USD）</div>

It can be concluded from Tab. 6, the cost of contract deployment and function execution is indeed low. Notably, the deployment of contract is a one-time charge and all the functions spend the fixed 'Gas' each time. It should be noted that the time period for storing data on-chain can be adjusted according to the actual practices. The time period for uploading data on-chain can be short, like storing records by day, for the purpose of higher security and availability, nevertheless, the cost will raise because the number of times of function execution will slightly increase. it is actually a trade-off between the cost and the security. Hence, we should minimize expenditures on the premise of ensuring safety in practice.

*5.2.3 Dynamic Authorization based on Trust Value*

In order to verify the rationality and feasibility of the user trust assessment method proposed in this study, this section artificially simulates behavioral records of user A, user B and user C generated on the Hadoop platform. For each user's behavior trajectory, only the RecordID, operations, time, and flag are extracted from off-chain database.

**The trust function.** In order to make the sigmoid function applicable to the scenario of user trust evaluation in Hadoop platform mentioned in this paper, we transform formula (5) into the following formula to reflect the dynamic change of trust value in a precise and parameterizable way. Specifically, the transformed function is denoted as follows.

$$ts_i = \frac{1}{1+e^{-(\frac{x-10y}{x+y})}}$$

(7)

In which, $x$ is the number of successful accesses, while $y$ represents the number of failed attempts, $ts_i$ is the corresponding trust value. In this paper, we set penalty factor $\beta = 10$. The effect of setting the penalty factor is that when the user commits evil in the system, it will have a great negative impact on its trust value, which can reflect the characteristics of rapid decline in trust value. Compared with the original sigmoid function, the function retains the characteristics of slow rise and trust value accumulation of the sigmoid function, meanwhile, it also improves in the following aspects:

1. The original sigmoid function: has a value of 0.5 when $x = y$, that is, the number of '*good*' times and the number of '*evils*' are equal, which is obviously unreasonable in the actual situation, while the value of improved function equals 0.5 when $x = 10y$.

2. The original sigmoid function ranges [0,1]. However, considering the actual situation, even if the user performs well all the time, it is impossible to assign the trust value to 1, while our transformed function reaches a plateau 0.731, when $x \gg y$ or $y = 0$, which is more realistic and reasonable.

3. The original sigmoid function, in the interval $x = [-\infty, -5], y \to 0; x = [5, +\infty], y \to 1$, which cannot handle input with more than tens digit, for instance, when $x = 100, y = 20$ ; $x = 100, y = 30$ ; $x = 100, y = 40$ , all the values approach 1, while the value of transformed function equals $z = 0.303; z = 0.177; z = 0.105$ separately(the trust value is declining with the failed access increasing).

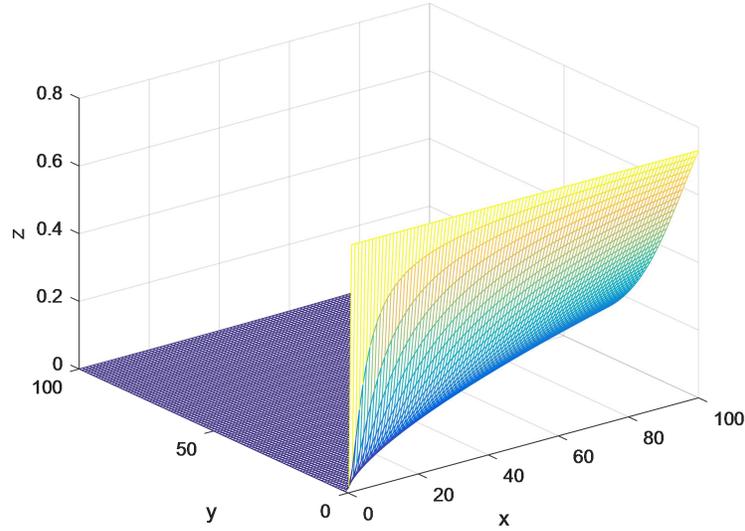

**Fig. 6.** The trust function

We can conclude from the above figure: when the *y* value is greater than that of *x*, that is, the number of '*evils*' is greater than the normal '*good*' times, the trust value is close to 0. When the '*good*' times is greater than the number of '*evils'*, as the difference between the two increases, the trust value increases slowly, but always less than 1.

To verify the validity of the trust function, we randomly generated three typical users' behavioral records for two months, and divided them into eight weeks.

**User A:** a user, who has relatively high initial trust value $T_{init} = 0.5$, and behaves well in the following two months (the number of successful visits is much greater than the number of failed visits).

**User B:** a user, whose initial trust value $T_{init} = 0.5$, but performs bad all the time in the subsequent visit.

**User C:** a user with relatively low initial trust value, $T_{init} = 0.2$, he behaves well in the first month, while poorly performs in the second month.

The artificial behavioral records of user A, user B and user C are seen as Tab. 7, and the fluctuation of the corresponding trust value calculated by the formula (7) is seen as Fig. 7.

**Tab. 7** User behavioral records

|   | Week1 | | Week2 | | Week3 | | Week4 | | Week5 | | Week6 | | Week7 | | Week8 | |
|---|---|---|---|---|---|---|---|---|---|---|---|---|---|---|---|---|
|   | suc | fail | suc | fail | suc | fail | suc | fail | suc | fail | suc | fail | suc | fail | suc | fail |
| A | 50 | 5 | 70 | 10 | 90 | 8 | 140 | 12 | 180 | 10 | 210 | 12 | 260 | 9 | 290 | 15 |
| B | 200 | 35 | 280 | 40 | 290 | 50 | 320 | 55 | 280 | 60 | 200 | 70 | 100 | 80 | 50 | 90 |
| C | 60 | 3 | 100 | 6 | 180 | 8 | 220 | 9 | 99 | 18 | 72 | 20 | 65 | 30 | 40 | 35 |

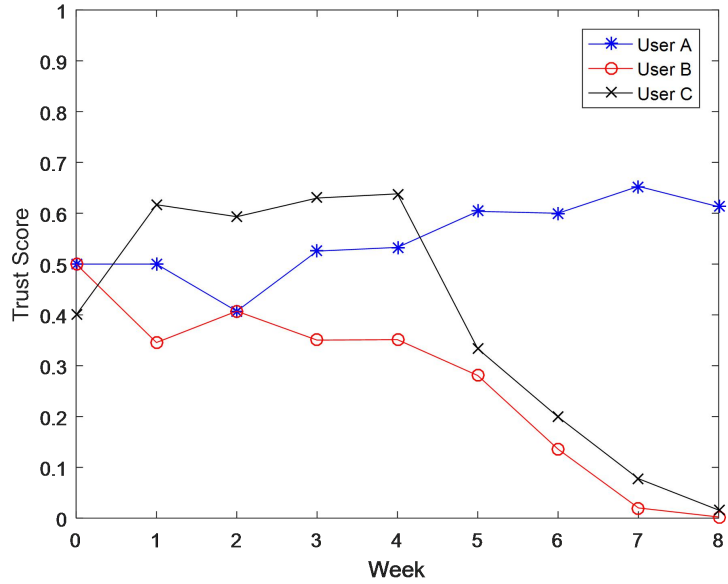

**Fig. 7.** Schematic diagram of the fluctuation of user trust value

**Comprehensive trust value calculation**

The data of each week are weighted according to the principle of "far small, near large". for ease of calculation, the weight of $1^{th}$, $2^{nd}$, $3^{th}$, $4^{th}$ week is assigned as $W = [0.1, 0.2, 0.3, 0.4]$, the impact of initial trust value is set as $\alpha = 0.5$. The comprehensive trust value of the first month as well as the second month are calculated through formula (6)(7).

$TS_{A_1} = 0.501, TS_{A_2} = 0.561$.

$TS_{B_1} = 0.431, TS_{B_2} = 0.247$

$TS_{C_1} = 0.512, TS_{C_2} = 0.307$

**Results Analysis**

**The trust value slowly rises and falls rapidly.** The initial trust value of user A is 0.5. Since the user A performs well in subsequent visits, the value increases to 0.561 after 2 months accumulation, which reflects the slow rise of the trust value. User B's initial trust value is 0.2. However, when the user behaves bad in subsequent visits, the trust value of the user quickly drops to close to 0, thus no resource can be accessed. The initial value of user C is 0.4, who performs well in the following 2 months, and the trust value slowly increased to around 0.65, while in the third month, the number of access failures increase, so its trust value declined rapidly. In summary, the trust function can reflect the slow rise and fast fall of the trust value.

**Dynamic authorization feature.** We assume that the resource access threshold of R1, R2, R3, R4 is 0.1，0.3，0.5，0.7 separately (which can be viewed from the resList maintained by the AP we mentioned earlier). AP adjusts permissions of users in real time according to the changes in their trust values, as shown in the following table.

**Tab. 8** Access changes with trust value

| Month / User | 0 | 1 | 2 |
|---|---|---|---|
| A | R1,R2,R3 | R1,R2,R3 | R1,R2,R3 |

| B | R1,R2,R3 | R1,R2 | R1 |
| C | R1,R2 | R1,R2,R3 | R1,R2 |

Initially, the default access rights of users A, B, and C are shown in the second column of the table above, and then the trust value is calculated according to the user's behavior in the platform, and their permissions are adjusted as seen the third column after one month. Similarly, after the second month, the permissions are dynamically adjusted according to the user trust value. It can be concluded that the proposed method can realize the function of adjusting permissions with the changing of the user's trust value.

## 6 Conclusion

In this paper, we have investigated the security issues of Hadoop platform and solved this problem by presenting the combination of ABAC-based and trust-based dynamic authorization scheme. We implement coarse-grained access control through the front-end access proxy as a centralized policy enforcement point, which frees back-end developers from the minutiae of dealing with authorization. For the sake of flexibility and dynamics, we modify the ABAC model to make it suitable for the big data platform, ABAC model makes access control authorization more reasonable by formulating a complete set of attribute policies and also taking environment factors into consideration. Moreover, to ensure fine-granularity, we also set an accessible trust threshold for each resource, and after satisfying the access policy of the access proxy, we need to determine whether the user has the right to access a resource according to the user's trust value. Using blockchain technology, the user behavioral record is stored in chronological order, which provides authenticity and tamper resistance to the evaluation of user trust. Meanwhile, we only use the attributes of 'successful access' and 'failures' as the input of the trust value function to calculate the user trust value. The implementation process is very simple and the model is highly scalable.

In summary, the dynamic authorization scheme offers a comparative advantage in addressing the secure access control issues with a combination of the coarse-grained, centralized authorization provided by the access proxy with ABAC and the fine-grained authorization implemented by the resource itself. Such collaboration can provide the maximum guarantee for the security of Hadoop platform.

**Conflicts of Interest:** The authors declare that they have no conflicts of interests to report regarding the present study.

## References

[1] CAICT. China Big Data Development Research Report[EB/OL]. (2018-04-26)[2020-06-04]. http://www.caict.ac.cn/kxyj/qwfb/ztbg/201804/P020180426332651074674.pdf.

[2] NIST. NIST Big Data Interoperability Framework: Volume 1, Definitions[EB/OL]. (2018-06-02)[ 2020-06-04]. https://bigdatawg.nist.gov/_uploadfiles/NIST.SP.1500-6r1.pdf.

[3] 360 CERT. MongoDB blackmail investigation report [EB/OL]. (2017-07-17)[ 2020-06-04]. https://cert.360.cn.

[4] S. Wang, "Analysis of the status of data leakage in transportation industry," *China Information Security*, vol. 000, no. 003, pp. 76-77, 2018.

[5]Infosecurity Group. Data Leak Exposes 267 Million Facebook Users[EB/OL]. (2019-12-20)[ 2020-06-04]. https://www.infosecurity-magazine.com/news/data-leak-exposes-267-million/.

[6] CVE. Apache Hadoop Vulnerability Statistics[EB/OL]. (2019-01-09)[2020-06-04]. https://www.cvedetails.com/product/22215/Apache-Hadoop.html?vendor_id=45.

[7] National information security standardization technical committee. Big Data Security Standardization White Paper [EB/OL]. (2018-04-16)[ 2020-06-04]. https://www.tc260.org.cn/upload/2018-04-16/1523808293220001658.pdf.

[8] S. Aditham and N. Ranganathan, "A system architecture for the detection of insider attacks in big data systems," *IEEE Transactions on Dependable and Secure Computing*, vol. 15, no. 6, pp. 974-987, 2017.

[9] P. P. Sharma, C. P. Navdeti, "Securing big data Hadoop: a review of security issues, threats and solution," *International Journal of Computer Science & Information Technolo*, vol. 5, no. 2, pp. 2126-2131, 2014.

[10] V. Shukla, "Hadoop security today & tomorrow," *Hortonworks Inc*, 2014.

[11] J. P. Anderson, "Computer security threat monitoring and surveillance," *Technical report*, James P. Anderson Co., April 1980. Contract 79F296400.

[12] M. R. Jam, L. M. Khanli, M. S. Javan and M. K. Akbari, "A survey on security of Hadoop," in *Proc. ICCKE*, Mashhad, Iran, pp. 716-721, 2014.

[13] Z. H. Wang, H. B. Pang, Z. B. Li, "Access control for Hadoop-based cloud computing," *Tsinghua Univ(Sci&Technol)*, vol. 54, no. 1, pp. 53-59, 2014.

[14] S. Liu, L. Tan, "New trust based access control model in Hadoop," *Computer Science*, vol. 41, no. 5, pp. 155-163, 2014.

[15] H. Y. Yang, L. X. Meng, "Hadoop cloud platform dynamic access control based on user behavior assessment," *Transactions of Beijing Institute of Technology*, vol. 37, no. 10, pp. 1031-1035, 2017.

[16] M. Kantarcioglu, "Securing big data: new access control challenges and approaches" in *Proc. SACMAT*, Toronto, ON, Canada, pp. 1-2, 2019.

[17] M. M. Shetty, D. H. Manjaiah and E. E. D. Hemdan, "Policy-based access control scheme for securing Hadoop ecosystem," in *Proc. Advances in Intelligent Systems and Computing*, pp. 167-176, 2019.

[18] J. H. Li, G. Zhao, S. Wang, Y. Z. Liu and J. Chen, "The research on personalized access control scheme based on user portrait," in *Proc. WHICEB*, pp. 66, 2019.

[19] Y. D. Wang, J. H. Yang, C. Xu, X. Ling and Y. Yang, "Survey on access control technologies for cloud computing," *Journal of Software*, vol. 26, no. 5, pp. 1129-1150, 2015.

[20] Y. K. Chen, X. L. Yin, W. L. Liu, "Access control model applicability for big data," *Information Security and Technology*, vol. 007, no. 007, pp. 3-5, 2016.

[21] Y. K. Chen, W. L. Liu, "Attribute-based access control model for Hadoop," *Journal of Henan Normal University(Natural Science Edition)*, vol. 44, no. 5, pp. 147-153, 2016.

[22] Z. Li, X. J. Xue, "Optimization of Cloud Service Trusted Evaluation Model Based on Sliding Window," *Computer Systems Applications*, vol. 28, no. 7, pp. 35-43, 2019.

[23] X. P. Liang, S. Shetty, D. Tosh, C. Kamhoua, K. Kwiat *et al.,* "Provchain: A blockchain-based data provenance architecture in cloud environment with enhanced privacy and availability," in *Proc CCGRID*, Madrid, Spain, pp. 468-477, 2017.

[24] D. Nguyen, J. Park, and R. Sandhu, "Dependency path patterns as the foundation of access control in provenance-aware systems," in *Proc TAPP*, 2012.

[25] L. Cittadini, B. Spear, B. Beyer and M. Saltonstall, "BeyondCorp part III the access proxy," *Usenix*, vol. 41, no. 4, 2016.

[26] H. Shafagh, L. Burkhalter, A. Hithnawi and S. Duquennoy, "Towards blockchain-based auditable storage and sharing of IoT data," in *Proc CCSW*, pp. 45-50, 2017.

[27] L. Q. Tian, C. Lin, "Evaluation mechanism for user behavior trust based on DSW," *Tsing hua Univ(Sci&Tech)*, vol. 50, no. 5, pp. 763-767, 2010.

[28] Apache Knox. http://knox.apache.org/.

[29] Ganache. https://www.trufflesuite.com/ganache.